# [1]Modelling Invasion Dynamics with Spatial Random-Fitness due to Microenvironment

Venkata S. K. Manem [1,*], Kamran Kaveh [1,**], Mohammad Kohandel [1]

and Siv Sivaloganathan [1, 2]

[1] Department of Applied Mathematics, University of Waterloo, Waterloo, Ontario, Canada

[2] Centre for Mathematical Medicine, Fields Institute for Research in Mathematical Sciences, Toronto, ON, Canada M5T 3J1

**Correspondence:** Venkata SK Manem (Email: vsmanem@uwaterloo.ca) or Mohammad Kohandel (Email: kohandel@uwaterloo.ca), Department of Applied Mathematics, University of Waterloo, 200 University Avenue West, Waterloo, Ontario. Email: vsmanem@uwaterloo.ca, Tel: +1-519-888-4567

**Abstract**

Numerous experimental studies have demonstrated that the microenvironment is a key regulator influencing the proliferative and migrative potentials of species. Spatial and temporal disturbances lead to adverse and hazardous microenvironments for cellular systems that is reflected in the phenotypic heterogeneity within the system. In this paper, we study the effect of microenvironment on the invasive capability of species, or mutants, on structured grids under the influence of site-dependent random proliferation in addition to a migration potential. We discuss both continuous and discrete fitness distributions. Our results suggest that the invasion probability is negatively correlated with the variance of fitness distribution of mutants (for both advantageous and neutral mutants) in the absence of migration of both types of cells. A similar behaviour is observed even in the presence of a random fitness distribution of host cells in the system with neutral fitness rate. In the case of a bimodal distribution, we observe zero invasion probability until the system reaches a (specific) proportion of advantageous phenotypes. Also, we find that the migrative potential amplifies the invasion probability as the variance of fitness of mutants increases in the system, which is the exact opposite in the absence of migration. Our computational framework captures the harsh micro-environmental conditions through quenched random fitness distributions and migration of cells, and our analysis shows that they play an important role in the invasion dynamics of several biological systems such as bacterial microhabitats, epithelial dysplasia, and metastasis. We believe that our results may lead to more experimental studies, which can in turn provide further insights into the role and impact of heterogeneous environments on invasion dynamics.

---

microenvironment, Fixation probability, Cellular Automata.

**Introduction**

The effect of spatial structure and heterogeneity is known to be of significant importance in evolutionary models, including evolutionary biological models and social networks (see [1-13] and references therein). The structure of the network might have subtle effects on the dynamical properties and other salient features of the model. One of the most important results obtained from these models is the fixation (also known as the invasion) probability. Invasion probability is defined as the probability that a species (eg. a mutant) can take over the whole population in the system. It is a measure of success of the selection process. Depending on the structure of the graph, the invasion probability can either be suppressed or enhanced and amplified. The question of how the structure of the network affects fixation has been the subject of much research. Maruyama showed that on regular graphs, the fixation probability is the same as on unstructured graphs [4-7]. Lieberman et al. [8] generalized this observation to a more general set of graphs known as isothermal graphs. This result is known as the isothermal theorem. Many authors have discussed the heterogeneity of the spatial structure and its impact on the fixation probability (see Manem et al. [9], Antal et al. [3], Sood et al. [4], Houchmandzadeh et al. [10]). Antal et al. [3] showed that upon introduction of randomness, in a scale-free random graph, the fixation probability is significantly suppressed. Other papers have suggested that particular configurations in fact enhance the fixation probability [8]. Recent work by Thalhausser et al. [12] and Manem et al. [9] on structured and unstructured meshes as well as random graphs showed that spatial structure influences the fixation probability to a greater extent for mutants with migration potential. Although a great deal of research has been devoted to the study of heterogeneous networks, less effort has been devoted to a study of heterogeneity, due to the spatial fitness distribution and environmental stress, and its effect on the invasion probability. Spatial variation of fitness, however, is a critical parameter in modelling biological and social systems as the fitness of species strongly depends on the microenvironmental parameters. For example, in models of bacterial growth, fitness can be a function of the spatial distribution of nutrients, and in social networks it can represent the geographical biases (see [11] for models of election). In viral and microbial evolutionary models, fitness can be a function of spatial distribution of drug concentration. It has been recently suggested that heterogeneity in environmental factor can increase the time to resistance in bacterial and viral dynamics by orders of magnitude ([14-16]). In the evolutionary dynamics of cancer, tumor progression is known to be strongly affected by the tumour microenvironment. In this paper, we focus our attention on modelling environment-induced fitnesses and their effect on the selection dynamics. We consider heterogeneity in environmental factors (such as drug concentration or resource values and microenvriomental parameters).

Another important aspect of environmental heterogeneity is its role in conferring phenotypic diversity. In principle accumulation of genetic alterations and/or changes in metabolic functions and adaptation to microenvironmental stresses (such as nutrients and growth factors) are considered to be key regulators of phenotypic heterogeneity [17-22]. This morphological diversity in the system plays a very important role in the evolution of many biological phenomena, i.e. it can act as a refuge for some species (i.e. permit co-existence of various species), or can facilitate fixation of a single species. Additionally, several microenvironmental conditions (such as scarcity of resources) encourage another genetic component, namely, migration of a species within a system. Thus, a relevant and interesting challenge is to understand the interplay of the migration potential and heterogeneous (fitness) distribution of mutants. However, as far as the authors are aware this problem has not been addressed in any evolutionary graph theoretical model of evolution to date. This scenario is addressed in the latter part of this work. We show that the interplay of spatially distributed fitness and the capability of individual species migration lead to non-trivial results for the invasion probability.

Quantifying the role of environment in evolutionary dynamics is a challenging task. Any mathematical model that attempts to do so should include heterogeneity in spatial structure including migration patterns as well as spatial heterogeneity of resources and/or hazardous environmental factors. Environmental stresses can affect both fitness of species as well as rate of mutation as a function of spatial position. For the sake of simplicity we avoid environmental effects on mutation rates and restrict ourselves to selection. We assume time scales much less than the time of appearance of (rare) advantaged mutants. Previous literature on spatially distributed fitness are limited and mainly from the fields of ecology and population genetics. Refs. [23][24] discussed meta-population models under diffusion approximation. The scope of these results are at best limited to weak selection in meta-population models. More recently, inspired by models of wealth distribution, [25] considered a model with background heterogeneity in fitness in an unstructured population and discussed fixation probability and time to fixation. The results are discussed for very small systems sizes without a spatial population structure using numerical simulations.

In this paper, we focus our attention on modelling the effect of environment on the selection process. We model the environment as a spatially distributed fitness function for both normal and mutant cells. We define a generalized evolutionary graph theoretical model that include spatial heterogeneity in the fitness distribution. The spatial variation in fitness is assumed to be random in space but constant in time. We address several challenging questions related to the interplay between spatial structure and different spatial fitness distributions and their effect on the fixation probability. We examine continuous distributions such as uniform and triangular besides discrete distributions like a bimodal distribution in the presence and absence of migration. We examine situations where a fraction of sites confer a fitness advantage for mutants and the rest of sites confer disadvantage and discuss the overall probability of fixation for a randomly placed mutant. We show that independent of the distribution of the fitness, heterogeneity is a suppressor of selection. Although the results in this work are applicable in a variety of evolutionary dynamical models, we focus our attention mainly on cancer progression models.

Another novel extension of our model is to introduce an independent migration potential to either of the phenotypes in the system. This is investigated by Thallhasuer et al in [12] for a uniform system. Two genetic factors are considered: replicative potential and cellular motility. The overall conclusion is that migration has a major impact on the probability of a single mutant cell's ability to invade an existing colony. This work was later extended by Manem et al. [9] who investigated the effect of migration on random and unstructured meshes. This is an effort to understand cancer progression on real tissue architectures. We show that migration potential of a mutant phenotype, in fact, can compensate for the critical suppression of selection due to fitness heterogeneity. In the context of evolutionary dynamics of cancer, due to increasing heterogeneity in microenvironment factors such as hypoxia and acidity, during later stages of cancer, this effect can be an evolutionary argument for the rise motile phenotypes (metastatic phenotype) inside a tumour.

**Materials and Methods**

Many authors have studied the spatial dynamics of cancer invasion using cellular automata models [26-29], and have progressively incorporated more complex cellular mechanisms. In these type of models, it is very likely that the effects of some forces on the underlying invasion dynamics are obscured by other dominant forces. In an attempt to understand and analyze the underlying dynamics, we consider a constant population model that incorporates the spatial structure of the environment as well as proliferation strength of the individuals. The fitness of each individual can also be affected by environmental factors that we include in the model. We assume a Moran process on a square lattice with death-birth updating. The Moran process was first described in [30], and subsequently used to study various scenarios concerning the selection dynamics of mutants in cancer biology [31-35]. Other modelling approaches related to environmental factors can be found in [36-40] (among many others).

Resident cells represent the population of either pre-cancerous cells or a dominant malignant phenotype inside the tumour which is being replaced by a more advantaged malignant mutant with more genetic alterations. We denote normal or resident cells as type A and mutant cells as type B. In a Moran process at each time step one individual is chosen randomly to die and another individual is chosen (proportional to its fitness) to replicate. This keeps the total population constant at each time step. In a death-birth updating, an individual is randomly chosen among the total population to die first and then another individual among the neighbouring connected sites is randomly chosen, with probability proportional to its fitness, to replicate. The neighbouring offspring replaces the dead cell with migration probability defined for the edge connecting the two neighbouring sites. Ref. [35] used this model to analyze the fixation probability of single hit and double hit mutants in one-dimension (1D). In particular, it was found that the probability of fixation of advantaged single mutants is lower compared to that in the space-free model. An extension of this 1D model is presented in [13]. In this paper, the authors derive approximate analytical results for the fixation probability for any degree-k regular graph.

Mutant and normal cells live on a regular graph. We assume a square lattice for simplicity. Each site $i$ can be occupied by only one individual. We assume that the fitness of each individual is position- or site-dependent. A mutant residing at site $i$ will have a fixed fitness of $r_{B,i}$ and a resident cell (normal cell) have fitness of $r_{A,i}$. Any constant term in $r_{B,i}$ and $r_{A,i}$ denote the intrinsic fitness of A or B phenotypes. Site-dependent parts, however, represent heterogeneity due to microenvironmental factors (for example, in the case of a tumour this would be due to vascular heterogeneity, hypoxic and acidic regions). Furthermore we assume that the fitness is randomly distributed. This is to account for the spatial heterogeneities and possible slow spatial fluctuations that can render some events as random. We assume that fluctuations in fitness due to changes in the microenvironment occur on much slower time scales than the selection time scales (time to fixation). We assume that rare (advantaged) mutants arise randomly at any site and to calculate fixation probability we average over all sites and all random configurations of fitness distributions. Later on in this paper, we will investigate the interplay between migration potential and the fitness distribution. We show that, in fact, when strong spatial heterogeneity might diminish chances of fixation for an advantaged mutant, the additional migratory potential of the mutant can overcome this barrier.

The probability for a normal or mutant cell to be chosen for death at site $i$ is assumed to be the same, $P_{A,i(\text{death})} = n_{A,i}/N$ and $P_{B,i(\text{death})} = n_{B,i}/N$. In other words, we assume cells have the same death rate. Here $n_{A,i}$ and $n_{B,i} = 1 - n_{A,i}$ denote population of A and B cells (0 or 1) at site $i$. A general model that considers variable death and birth rates for both phenotypes is discussed in [13]. Similarly, the probability that a A (B) cell is chosen for reproduction is:

$$P_{A,i(\text{div})} = \frac{\sum_j w_{ij} \, r_{A,j} \, n_{A,j}}{\sum_k w_{ik} \, (r_{B,k} \, n_{B,k} + r_{A,k} \, n_{A,k})} \quad , \quad P_{B,i(\text{div})} = \frac{\sum_j w_{ij} \, r_{B,j} \, n_{B,j}}{\sum_k w_{ik} \, (r_{B,k} \, n_{B,k} + r_{A,k} \, n_{A,k})}$$

The migration matrix $w_{ij}$ is defined to be one over the number of neighbours (1/4 for square lattice, von-Neuman), if $i$ and $j$ are neighbouring sites and is zero otherwise. The sums over indices $k$ and $j$ are effectively over the neighbouring sites. In other words, at each site a cell is chosen with probability $1/N$ to die and the neighbouring occupants offsprings replace the empty site with rates proportional to their fitness which depend on their positions in the graph.

To include the possible migrative potential of the two phenotypes, a second generalization of this model is introduced in [12]. By this migrative potential we mean migration that does not accompany a proliferation event and merely represents a motility potential of the phenotype. The update rules for the model are as follows: Suppose $m_A$ and $m_B$ are the migrative rates (cellular motility) of type A and B cells respectively. Now at each time step a cell is chosen randomly to die (say at site $i$). Then one of the following four events might occur for a neighbouring cell: (1) a neighbouring A cell divides and the offspring moves to site $i$, (2) a neighbouring A cell migrates to site $i$, (3) a neighbouring B cell divides and the offspring moves to site $i$, or (4) a neighbouring B cell migrates to site $i$. If a reproduction event of either A or B occurs, then the update is complete and the process is repeated again. Suppose that a migration event of either A or B occurs, then the empty spot is occupied by a migratory cell, leaving another empty spot behind. Again, a new elementary event is considered till the empty spot is filled (i.e., until the occurrence of a birth event). The lattice is always filled up at the end of each update following the Moran process assumption that the whole population is constant at every iteration. The probabilities of each proliferation or migration event are given by:

$$P_{A,i(\text{div})} = \frac{\sum_j w_{ij} r_{A,j} n_{A,j}}{N_{r,i}} \quad , \quad P_{B,i(\text{div})} = \frac{\sum_j w_{ij} r_{B,j} n_{B,j}}{N_{r,i}}$$

$$P_{A,i(mig)} = \frac{\sum_j m_A \, n_{A,j}}{N_{r,i}} \quad , \quad P_{B,i(mig)} = \frac{\sum_j m_B \, n_{B,j}}{N_{r,i}}$$

where $N_{r,i} = \sum_k w_{ik} [(r_{B,k} + m_B) n_{B,k} + (r_{A,k} + m_A) n_{A,k}]$. Notice that $n_{B,j} = 1 - n_{A,j}$ as before. We assume that motilities, denoted by $m_A$ and $m_B$ for normal and mutant cells, are independent of the sites and uniform all over the system.

In the following we focus on a square lattice with reflecting boundary conditions. First we discuss the effect of the random distribution of fitness while assuming no motility (i.e. $m_A = m_B = 0$). The question is if inclusion of the heterogeneity in the fitness can suppress or amplify selection. This is represented by the value of the fixation probability as a function of the fitness distribution width. First we assume that the fitness of the normal cells is not affected by the microenvironmental factors, $r_{A,i} = 1$ and that only mutants have a site-dependent fitness $r_{B,i}$. We treat $r_{B,i}$ as random variable with a given probability distribution function. We first consider uniform distribution for mutant fitness with distribution width $\Delta$. Next we include random fitness variations in *both* normal/resident cells and mutant cells. To evaluate the fixation probability, we perform exact stochastic simulations over a $21 \times 21$ square lattice and average over the initial positions of mutant cells *and* also the random configurations of fitness distribution.

The critical question in this experiment is to analyze if the selection dynamics depends on the shape of the fitness distribution. We assume fitness distributions where the mean value of mutant fitness and normal cell fitness coincide. In other words, we also consider distributions with skewness where mean and median of the distribution are not the same. As an example of such distributions we will examine a triangular distribution. It should be noted that the fixation probability is time independent quantity, so that in our work, an iteration is defined as the system reaching one of its absorbing states (i.e. the system is either filled with type A cells or type B cells). A set comprises a number of iterations in order to obtain a statistical average of the fixation probability. We use the death-birth (or later death-birth/migration) updating algorithm outlined above and run 10000 iterations for each fixed random configuration of fitness values. We repeat this for several configurations of fitnesses and average over the results. The error bars in all the figures represent the standard deviation from the average. The error bars denote the uncertainty in the average over both stochastic iterations and random configurations of fitnesses.

To consider highly heterogeneous systems, we consider simple *discrete* probability distributions for fitness such as a Bernoulli distribution (bimodal distribution). A given percentage $x$ of sites confer a fitness advantage to mutant population where rest of sites, $1 - x$ confer disadvantage or no advantage. For low densities of advantaged sites the fixation probability is zero and close to a percolation limit it begin increasing. In weak selection it follows a Moran result in unstructured populations and deviates from Moran for higher average fitness. The analytical investigation of the above cases is the subject of future work.

Finally, we consider the effect of the migrative potential in the presence of a discrete distribution

(bimodal distribution) on the mutant fitness. The interplay between cellular motility and the environment-induced fitness distribution is very interesting. Our findings suggest that while harsh microenvironmental factors can critically suppress selection for a mutant phenotype, the gain of motility potentials can compensate for this. Even in presence of strong fitness heterogeneity, fitness waves can travel all through the system for strong enough motility.

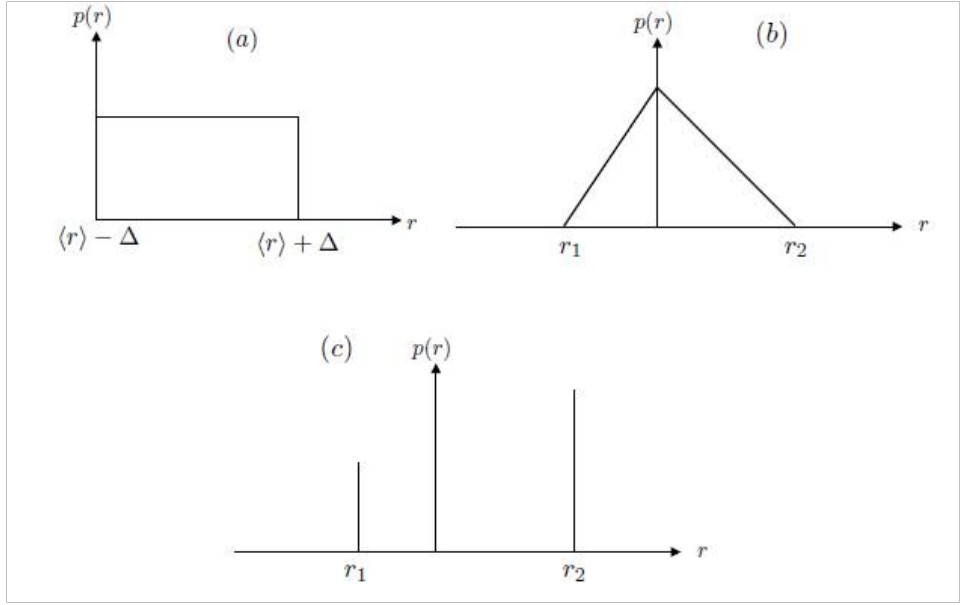

**Figure 1:** Different forms of random fitness distributions: a) Uniform b) triangular c) bimodal.

**Results**

In this section, we investigate the effect of randomness of mutants in a system, i.e. when one or both of the two phenotypes have random fitness from a given distribution. Figure 1 displays different types of fitness distributions chosen for the numerical experiments: two continuous distributions (uniform and triangular, $D = \{0.9, 1.25\}$ and $D = \{0.75, 1.1\}$) and a discrete distribution (bimodal). $D$ indicates maximum and minimum allowed values for fitnesses, and we denotes $\Delta$ by the width of the distribution. We denote the average over random configurations of fitness by $\langle r \rangle$. A uniform distribution is defined as fitnesses in the range $\langle r \rangle - \Delta$ to $\langle r \rangle + \Delta$ having the same probability of being assigned for any site in the system. A triangular distribution is similar to uniform distribution but has the property that its mean and median might not coincide. In other words if the range of values for fitness are $r_0 - \Delta$ and $r_0 + \Delta$ the average $\langle r \rangle$ is not necessarily $r_0$. We examine this case when the median $r_0$ for mutant fitness equals the normal cell fitness and see if skewness in the probability distribution can confer an overall advantage or disadvantage to mutants. Finally a bimodal distribution is a simple Bernoulli distribution where at each site cells can either have $r_1$ or $r_2$ fitness values. The fraction of sites with fitness $r_2$ is denoted by $x$.

*Random Fitness of Mutants only*

In this section, we investigate the effect of variation of the fitness of the mutant cells (B cells) on the invasion probability. This is carried out by assuming a fixed fitness for type A cells, which does not depend on the environment and assumed constant. Also, we take the fitness of type A cells as the reference fitness (by assuming it to be equal to 1). As indicated in Figure 1(a) the fitness of B cells in this case, follows a uniform distribution where $\langle r_B \rangle - \Delta < r < \langle r_B \rangle + \Delta$. For this, we construct a regular grid consisting of $21\times 21 = 441$ elements with reflecting boundary conditions, and fix the fitness rate of A cells, i.e. $r_A = 1$. We then randomly generate fitness rates for B cells at each nodal point on the lattice such that the average value is $\langle r_B \rangle = 1.5$ for advantageous mutants and fix this fitness matrix for all the iterations. We fill the lattice with type A cells at each nodal point on the grid, and randomly place a mutant in the system (by deleting an A cell). So, the system is now comprised of $N-1$ type A cells and 1 type B cell. The algorithm outlined in the previous section is applied with reflecting boundary conditions until the system reaches one of its absorbing states. Figure 2 shows the effect of varying the distribution width of $r_B$ on the invasion probability for advantageous mutants.

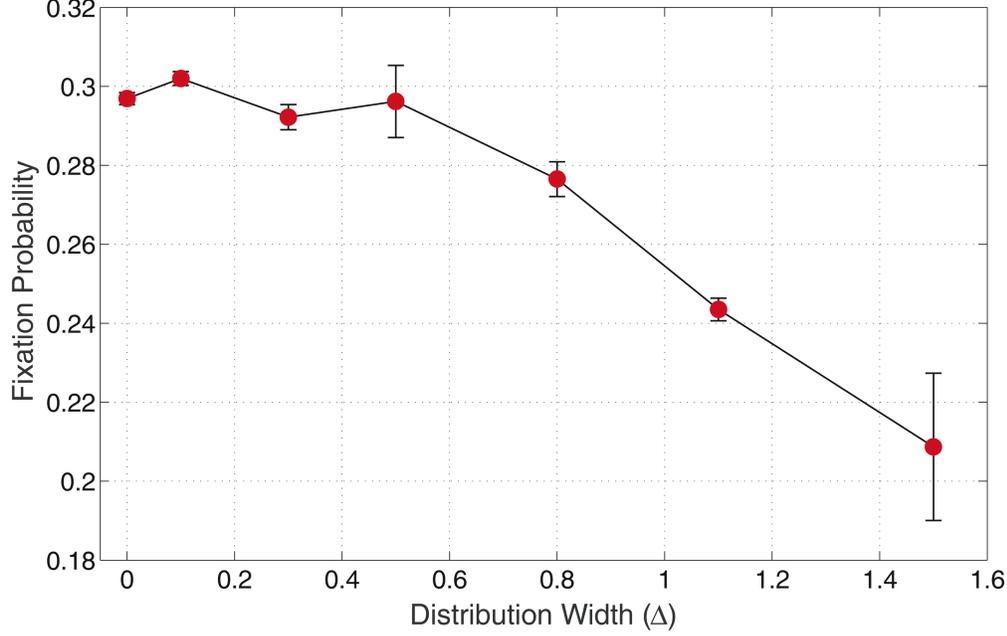

**Figure 2:** Invasion probability of mutants as a function of varying distribution width of fitness $r_B$. Parameters: Lattice size= $21\times 21$, $r_A = 1.0$, $\langle r_B \rangle = 1.5$, $m_A = m_B = 0$ (Spatial model in the absence of migration of cells).

From Figure 2, we observe that the invasion probability negatively correlates with the width of the distribution, i.e., as the width of the fitness distribution of type B cells $\Delta$, increases the invasion probability decreases for advantageous mutants. In Figure 2, we do not observe a huge impact on the invasion probability till the value of $\Delta = 0.4 - 0.5$. Once the $\Delta$ value increases across the threshold of 0.4-0.5, i.e. relatively small width, we notice a decline in the invasion probability. At the lower end of the distribution, i.e. $\langle r_B \rangle - \Delta$ reaches $r_A = 1$ one can see that the fixation probability begins to decrease at a much more significant pace. This indicates that, as the number of sites where B cells are *disadvantaged* gets introduced in the system, the fixation probability decreases more significantly. (Notice that the drop has almost a linear form, but, this may well be an artifact of the finite size of the system used in the simulation.)

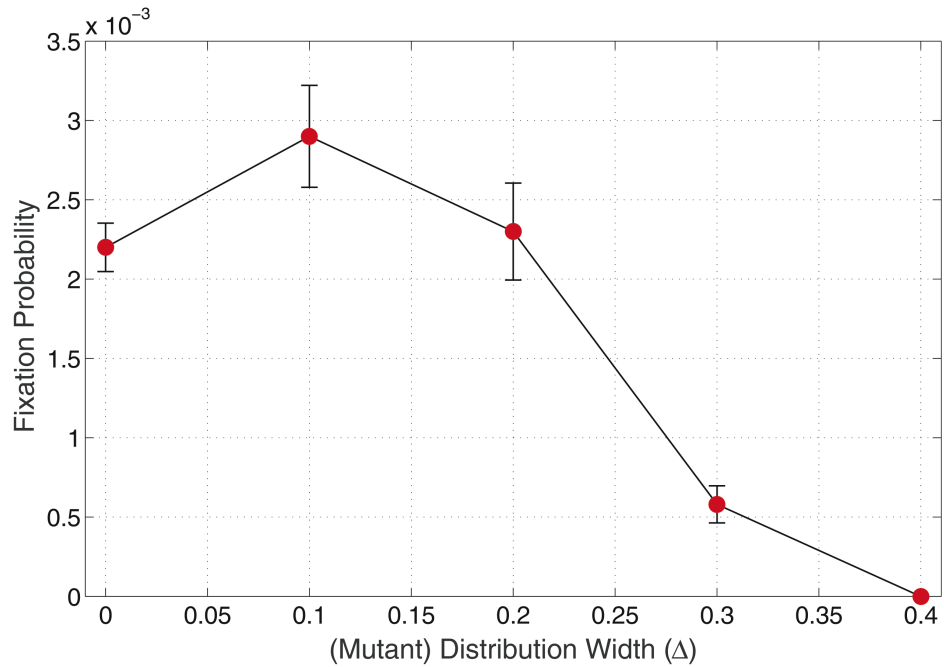

**Figure 3:** Invasion probability of mutants as a function of varying distribution width of fitness $r_B$. Parameters: Lattice size=$21 \times 21$, $r_A = 1.0$, $\langle r_B \rangle = 1.0$, $m_A = m_B = 0$ (Spatial model in the absence of migration of cells).

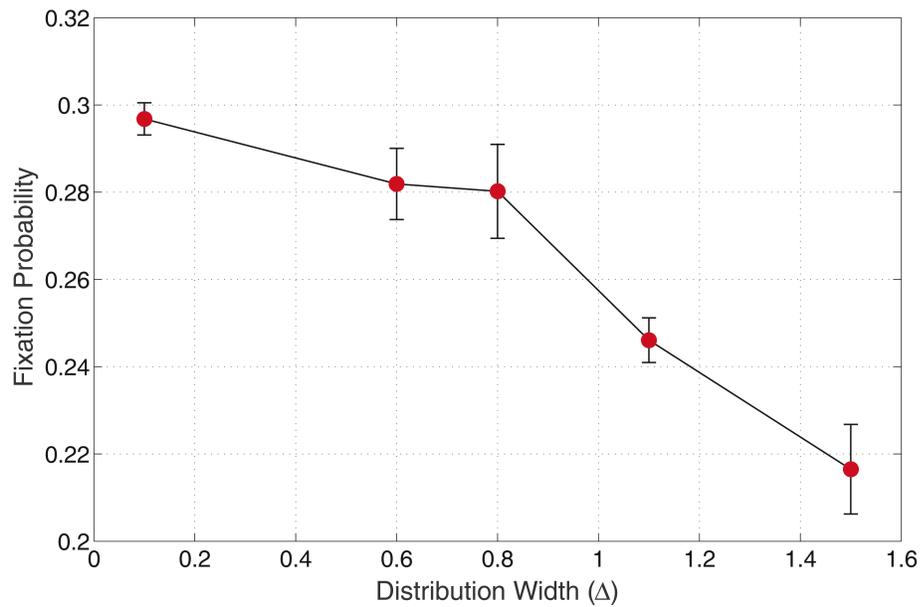

**Figure 4:** Invasion probability of mutants as a function of varying distribution width of fitness $r_B$. Parameters: Lattice size= $21 \times 21$, $\langle r_A \rangle = 1.0$ $\langle r_B \rangle = 1.5$, $m_A = m_B = 0$ (Spatial model in the absence of migration of cells).

For a high variance of the distribution width, we can intuitively imagine the existence of various types of B cells (with different fitness rates) which are competing against each other, and also against type A cells in the system. Due to the intense competition between different fitnesses type B cells, and also against type A cells, it becomes hard for mutants to survive, resulting in a sharp decline in the invasion probability for higher variance of $r_B$. Figure 3 displays a plot of varying distribution width of $r_B$ on the invasion probability for neutral mutants.

For zero distribution width, we obtain a value of the invasion probability approximately equal to the value of the space free fixation probability $\rho = 1/N = 0.0023$. Although the simulations were run for a large number of iterations, we do observe a slight increase in the invasion probability for $\Delta \approx 0.1, 0.2$ which can be attributed to numerical errors in the simulations.

*Random Fitness of Host Cells and Mutants*

In this section, we investigate the effect of random fitness distributions for *both* types A and B cells. We chose the fitness distribution width for type A cells to be $D = \{0.9, 1.1\}$ with the average value chosen to be $r_A = 1$, and varied the distributions for type B cells. Figure 4 displays a plot of invasion probability against the variance of $r_B$ for advantageous mutants whose average value is 1.5. We notice that as the variance of type B cells increases, the invasion probability decreases. Although, it is interesting to note that from Figures 2 and 4, for a $\Delta$ value of 1.5, the values of $\rho = 0.2087$ and $\rho = 0.2165$ for non-random and random fitness rates of type A cells respectively. For a distribution width of 0.1 away from the average fitness value of A cells, we do not observe a huge change in the invasion probability when compared to the non-random fitness rate of A cells (see Figure 2).

Probably, the effect of random fitness rates of type A cells on the invasion probability might be clearly observed if the width is increased to a larger value. From our results, we observe that a small change in the fitness rate distribution of host cells does not have much impact on the invasion probability.

Another interesting observation is that, even though one intuitively expects that as the distribution of A cells broadens towards more advantageous states and as the distribution of B cells broadens towards less advantageous states, the point at which we observe a change in regime (bending of the invasion probability curve) should be when the two distributions begin to overlap. As a matter of fact, this is not the case, and illustrates the point that the undergoing mechanisms are much more sophisticated and non-trivial.

Figure 5 displays a plot of the invasion probability against the variance of $r_B$ for neutral mutants whose average value is 1.0. As in the previous cases, we observe a negative correlation of the distribution width of mutants with the invasion probability even in the presence of random fitness of host cells. It is also interesting to compare Figures 3 and 5, for a $\Delta$ value of 0.3. We notice that the presence of random fitness of host cells pushes the invasion probability (see Figure 5) to a smaller value compared to that obtained in the absence of randomness of host cells (see Figure 3). This can be attributed to the competition between various types of host cells (with different fitnesses) as well as with other mutants, resulting in a decline in the invasion probability. This is again due to the fact that the fraction of disadvantaged (i.e. $r_B < r_A$) type B cells increases in the system.

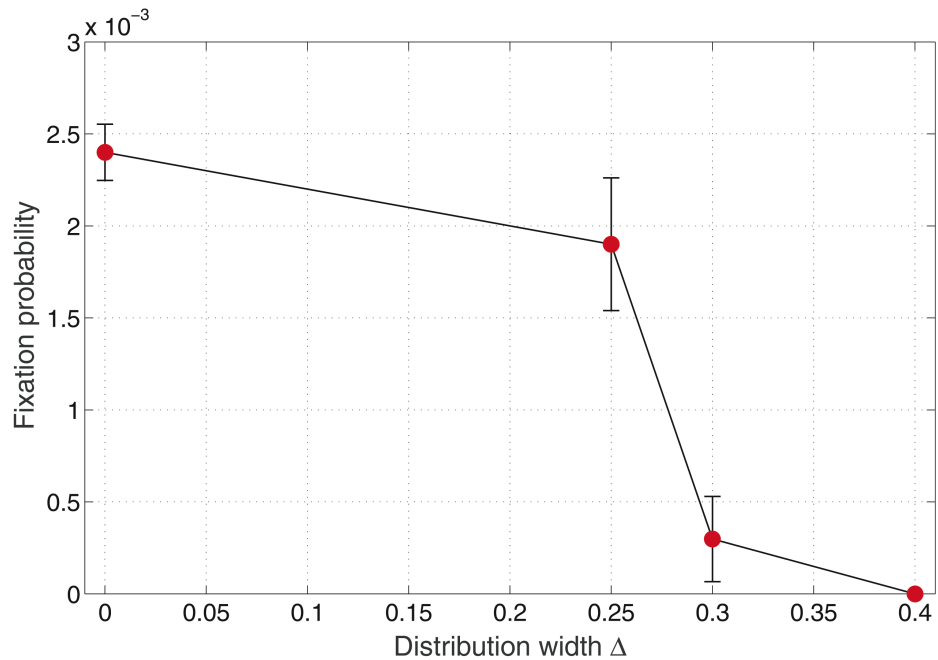

**Figure 5:** Invasion probability of mutants as a function of varying distribution width of fitness $r_B$. Parameters: Lattice size=$21 \times 21$, $\langle r_A \rangle = 1.0$ $\langle r_B \rangle = 1.0$, $m_A = m_B = 0$(Spatial model in the absence of migration of cells).

*Random Fitness with Neutral Average*

In this section, we choose the fitness distribution width of neutral mutants to be a triangular distribution (i.e. a simple distribution with skewness). The fitness of type A cells is fixed to be $r_A = 1$. The choice of a triangular distribution lets us fix the mean value of both A and B cells to be the same (i.e. $r_A = r_B = 1$), while the median value of B can be greater than unity. We tested two scenarios to investigate if a neutral mutant can be pushed to an advantageous or a disadvantageous state for 2 sets of triangular distributions. We choose the triangular distributions in such a way that the average value of $r_B$ equals that of $r_A$ (although numerically it is close to but not exactly equal to 1.0).

The values of $\rho$ for the two types of triangular distributions (called $D_1$ and $D_2$) are close to zero and 0.045767 respectively. This clearly indicates that a neutral mutant can be pushed either to an advantageous mutant state (if $D_1 = \{0.9, 1.25\}$), or, to a disadvantageous mutant (if $D_2 = \{0.75, 1.1\}$). The value of $\rho$ in a space free system is $\rho = 1/N = 0.0023$, and, on a structured grid $\rho = 0.00224$. However, if the system has fitness rates from a triangular distribution $D_2$, then the invasion probability is increased by almost 95%. At the same time, if the fitness distribution is from $D_1$, then the mutant loses its neutral drift and becomes a disadvantaged mutant, leading to zero invasion probability. As can be seen the effects are minor, but still observable in our simulations.

Biologically, it is possible that fitness distributions may not always be uniformly distributed in the tissue. From our computational analysis, neutral mutants cannot be pushed to become advantageous mutants for a uniform fitness distribution. However, if the distribution is slightly changed to triangular, (i.e., the resources and nutrients vary according to the triangular distribution), then a neutral mutant can turn into an slightly advantageous mutant. This is one of the plausible reasons why we might (or, might not) observe the invasion mechanism under harsh micro-environmental conditions for *neutral* mutants.

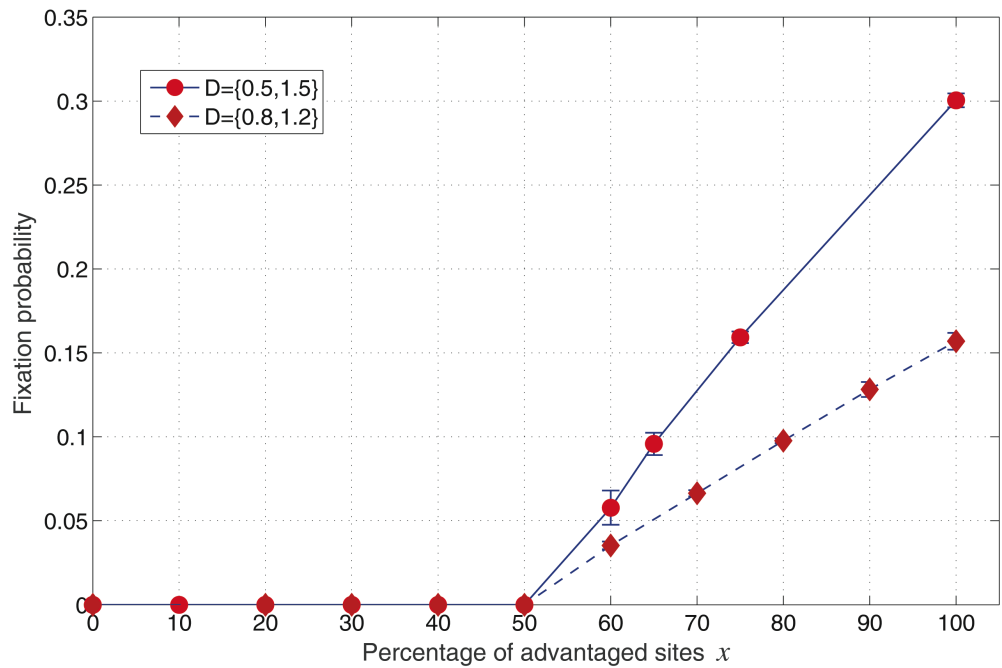

**Figure 6:** Invasion Probability against the percentage of advantageous mutants. Parameters: Lattice size = $21 \times 21$, $r_A = 1.0$, $m_A = m_B = 0$ (Spatial model in the absence of migration of cells).

### *Strong Heterogeneity and Localization*

In this section, we investigate a discrete distribution of fitness values. This can be a good model for strong heterogeneities in the system. Also it can model many binary changes in phenotype that are influenced by environment. For example, cells can transition into a quiescent state in response to lack of nutrients, or due to a harsher environment (hypoxia, acidity, drug delivery) thus reducing cell cycle times and fitness significantly. For this purpose, we choose the fitness distribution of mutants from a bimodal distribution. We choose $r_B$ belongs to the set $r_B \in \{r_1, r_2\} = \{0.5, 1.5\}$ and $r_B \in \{r_1, r_2\} = \{0.8, 1.2\}$ (with average 1.0), and fix the fitness of host type cells to be equal to one. We construct the fitness vector for mutants in such a way that the system is comprised of $x$ percent ($0 < x < 1$) mutants with advantageous fitness rate, $r_2$, and the rest with disadvantageous fitness rate, $r_1$. The result of stochastic simulations for fixation probability is plotted in Figure 6. It displays invasion probability against the percentage of mutant phenotypes with advantageous fitness rate. The percentage $x$ is a measure of environmental diversity in the bimodal distribution. Notice that $x$ is a measure of the width of the distribution as well.

As seen from Figure 6, the invasion probability for the mutants ($r_2 = 1.5$, or $r_2 = 1.2$) is zero until the bimodal distribution reaches a point where more than half of the points in the lattice confer advantageous fitness. This can be understood by assuming a uniform and even distribution of the two fitness values (say, $r_B = 0.5$, or $r_B = 1.5$) on a square lattice (i.e. checker-board distribution). Now every advantaged mutant is merely one lattice site away from a fellow advantaged mutant, and upon one division event this gap is filled which gives (almost) connected regions of advantage to the mutants. A much lower abundance of sites with advantaged fitness values leads to separated islands inside which, the average value of fitness would confer a fitness advantage for the mutants there. However due to the distance between them, the chance to capture the whole system would be extremely small. In other words, a successful invasion of the system is possible if all different islands with advantaged fitness sites are tightly connected. Due to the construction of the death-birth model the regions with on-site advantaged fitness can be one cell away from each other and still lead to a successful invasion. The fact that all the regions with advantageous fitness, (i.e. $r_2$ in Figure 1c) should not be separated or found far from each other is illustrated in Figure 7a. The ideal checkerboard distribution where every other site has an advantageous value of fitness for B cells is depicted in Figure 7b.

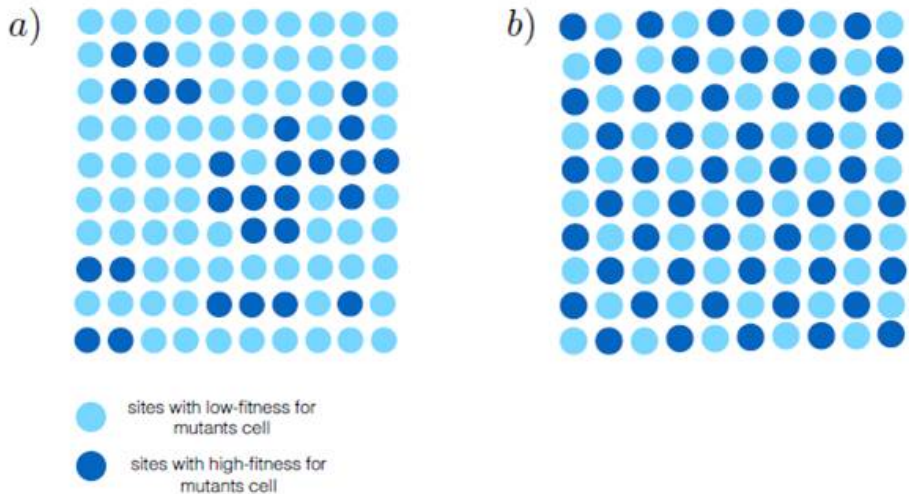

**Figure 7:** (a) Random distribution of two types of sites with high and low fitness's for mutant cells. (b) Evenly distributed high and low fitness sites with equal ratios (ideal case). This is the onset of successful fixation.

In the case of a bimodal distribution we can compare our results with that of a uniform system. In a uniform system, the fixation probability can be replaced by the following Moran model formula:

$$\rho_{bimodal} \approx 1 - \frac{1}{\langle r \rangle}, (\delta r << 1)$$

$$\approx 1 - \frac{1}{r_2 x + r_1 (1-x)}.$$

where $r_1$ and $r_2$ are the values of fitness, and $x$ is the fraction of $r_2$ sites, i.e. disadvantageous sites. Figure 8 displays the comparison of invasion probabilities between analytical approximation and simulated results. As can be seen for the case of $r_1 = 0.5$ and $r_2 = 1.5$ where $\langle r_B \rangle = 1.5 \neq r_A$, the above simple approximation breaks down and the difference in error increases. We have also simulated for the scenario: $r_1 = 0.2$ and $r_2 = 2.8$ (not shown here), and in this case we observed that the analytical approximation is no longer valid, and the error increases between simulated and analytical value. Detailed derivation of analytical solutions for a simple model such as this is very tedious and is left for future work. However the main observation from our simulations remains true, that in weak-heterogeneities and in the weak-selection limit, the fixation probability follows that of a uniform system (Moran result).

*Interplay of Migration and Fitness Heterogeneity*

In this section, we investigate the effect of the migrative potential of mutants on the invasion probability for a random fitness distribution width of type B cells. This experiment is carried out in order to try and ascertain if migration reduces or amplifies the invasion probability. The migrative potential of host cells is assumed to be $m_A = 0$, while that of mutants is chosen to be $m_B = 1$ (low migrative potential) and $m_B = 5$ (high migration potential). The algorithm outlined previously is carried out with reflecting boundary conditions, until the system reaches one of its absorbing states. Figure 9 displays the plot between the variance of distribution width of mutants against the fixation probability in the presence of migration.

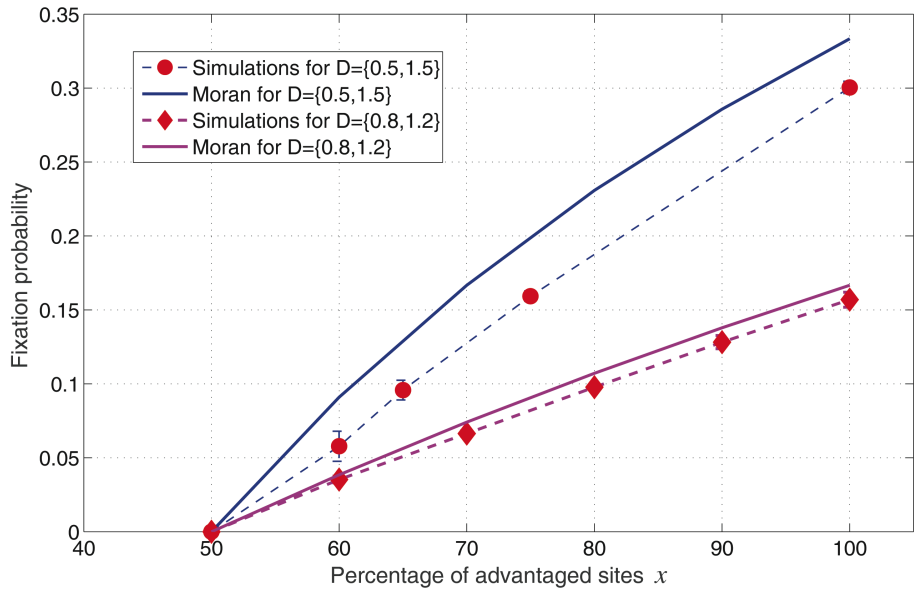

**Figure 8:** Comparison of invasion probabilities- Analytical and Simulated values for bimodal distributions of variable width (Spatial model in the absence of migration of cells).

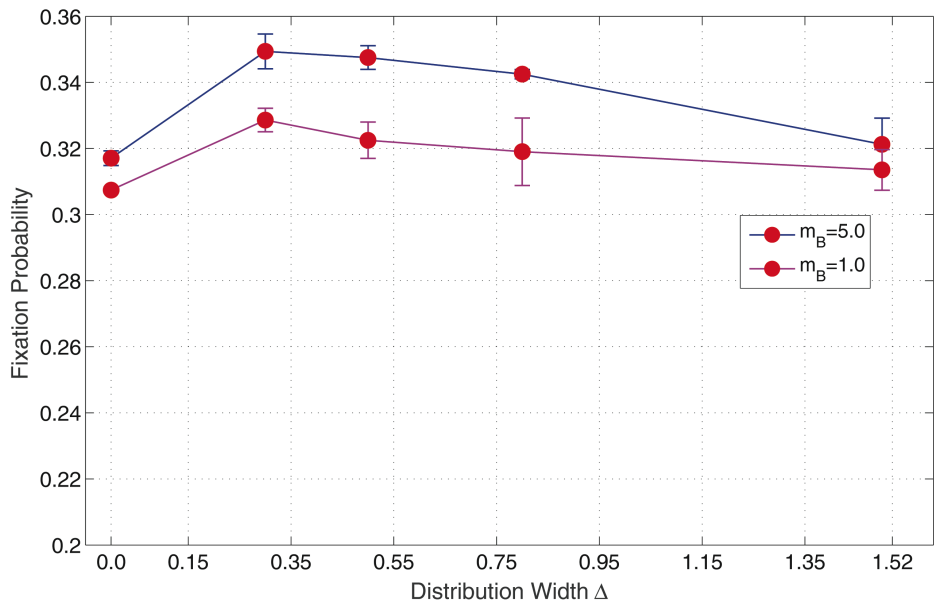

**Figure 9:** Invasion probability of mutants as a function of varying distribution width of fitness $r_B$. Parameters: Lattice size=$21 \times 21$, $r_A = 1.0$, $\langle r_B \rangle = 1.5$, $m_A = 0$ (Spatial model in the presence of migration of mutants only).

From our computational analysis, it is clear that the migration potential of mutants amplifies the invasion probability. In Figure 2, we notice that in the absence of migration, the invasion probability decreases as the variance increases. This scenario is quite the opposite of Figure 9, wherein the invasion probability increases in the presence of migration. Invasion probability increases as the variance increases, but almost stays constant for larger variances. As indicated in Figure 10 this increase can be interpreted to be the result of the migration potential of mutant cells allowing the isolated, distant high fitness regions the ability to communicate with other types and thus result in a higher invasion probability, even though these are smaller percentages of high fitness (advantaged) sites. One can also notice that the $\delta$ value for which the behaviour of the invasion probability begins to change is around the same value as that obtained without migration.

A similar experiment is carried out for neutral mutants, by varying the distribution width of mutants in the presence of migration. Figure 11 displays a plot of invasion probability against the variance of fitness distribution of type B cells. A similar behaviour is observed as in the previous plot. The invasion probability is amplified in the presence of migration even for neutral mutants, and stays almost constant for an increasing distribution width.

**Discussion**

We have presented a computational framework to understand the effect of the microenvironment on the invasion probability. An obscure biological phenomenon that is not well understood is phenotypic heterogeneity in a system. Several experimental studies have suggested that the selection dynamics is quite different in a heterogeneous microenvironment compared to a homogeneous one. In a homogeneous medium there is always a possibility of a single clone outgrowth, while in a highly heterogeneous environment, multi-clones can co-exist in various (non-connected) niches within a system. Additionally, this complex behaviour can be altered by the introduction of a migration potential for the individuals in the system. Harsh microenvironmental conditions are defined as species having varying fitness rates at each point in the system (which are independent of time). We have considered cellular automata rules based on biological assumptions, and not imposed them in order to obtain interesting salient features of the model. Thus, we have developed a simplified model that has two types of phenotypes in the system, namely host cells and mutants. The reproductive rates of both host cells and mutants might differ due to several microenvironmental conditions. For example, the division rate of a phenotype might be modulated due to the rigidity of the host tissue, or due to lack of nutrients at that position. Our computational model encapsulates these into a spatially distributed random fitness of mutants as well as of the host cells in the presence and absence of migration, and shows the impact of this on the invasion probability.

The effect of random fitness distribution widths of advantageous and neutral mutants (in the absence of migration) on the invasion probability is examined. From our simulations, we observe that the invasion probability decreases as the variance of the fitness distribution increases. A fitness distribution with zero width (which can be characterized as a well oxygenated system) gives rise to a higher invasion probability (which can be understood intuitively) compared to a non-oxygenated system. However, as the variance of the fitness distribution increases (i.e. under harsh microenvironmental conditions), the invasion probability decreases. This can arise as a result of the intense competition between various mutants (that emerge due to changes in the environment), and also competition with the host cells to invade the system.

We also focused on the impact of the invasion probability, due to a random fitness distribution of both the host cells and also mutants, on a migration free system. As noted in the previous scenario, the invasion probability negatively correlates with the distribution width. It is also interesting to note that there is no significant difference between the invasion probabilities in the presence and absence of random fitness of host cells. This can be attributed to the less randomness (with width one away from the average value) of host cells in the system, which behave in quite a similar fashion to a distribution of zero width. The increase in the variance of the distribution can be interpreted as the system having varying fitness rates at each point. Several mutants (with different fitness rates) arise in the system and compete against each other, and also with the host cells, to take over the whole tissue. Increasing the variance implies an increase in the number of different mutant phenotypes in the system competing against each other to take over the whole tissue, and thus results in a decline in the invasion probability.

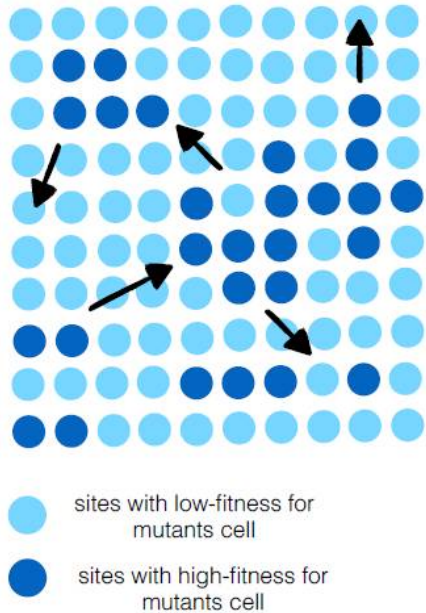

**Figure 10:** In the presence of migration the previously isolated regions with advantaged fitness for B cells can communicate due to their migration potential (indicated with arrows).

Additionally, in this paper, we have investigated another scenario by choosing the fitness distribution from a bimodal distribution $r_B = \{r_1, r_2\} = \{0.5, 1.5\}$ to capture the invasion feature of mutants. It is interesting to note that the invasion probability is zero until the system reaches a particular number of advantageous mutants. The invasion probability is zero even when the system is filled with 50% of advantageous mutants. This zero invasion probability can be attributed to the competition between the 50% of advantageous mutants and remaining 50% of disadvantageous mutants, as well as against the host cells in the system. This suggests that under moderate (non-harsh) microenvironmental conditions, we observe the invasion probability to be zero until the system is dominated by advantageous mutants.

Also, the effect of random fitness distribution widths (of mutants) on the invasion probability, in the presence of high migration potential of mutants, has been investigated. From our analysis, we observe that migration amplifies the invasion probability as the variance increases for both advantageous and neutral mutants. In a migration free system, the invasion probability is suppressed as the cells do not move in the system, due to which there is a higher chance of mutant extinction. However in the presence of migration, the extinction probability of mutants is small due to cell movement within the system. Hence, we observe an amplification of the invasion probability for both advantageous and neutral mutants.

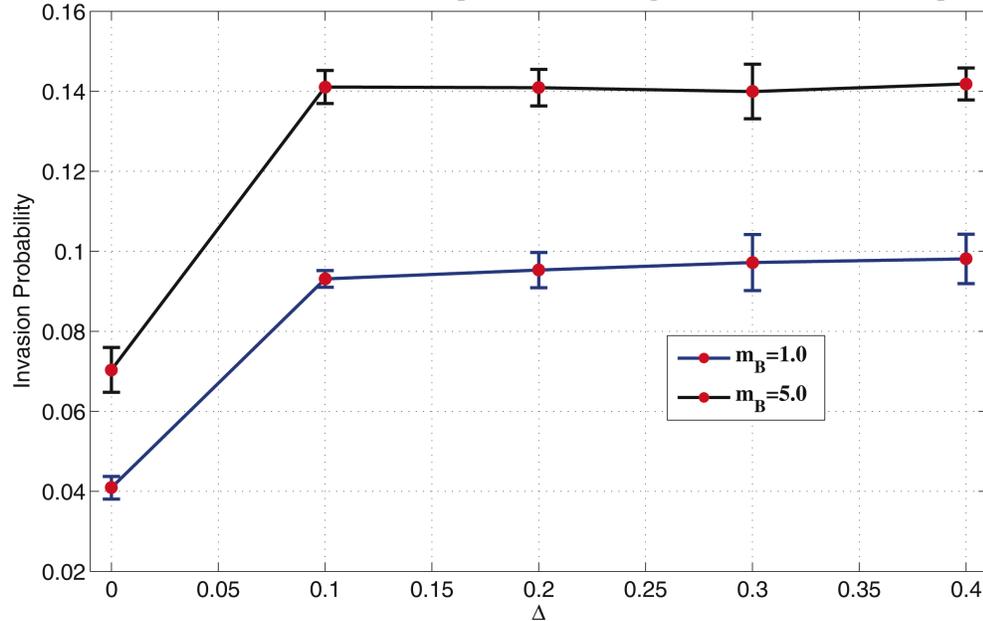

**Figure 11:** Invasion probability of mutants as a function of varying distribution width of fitness $r_B$. Parameters: Lattice size= $21\times 21$, $r_A = 1.0$, $\langle r_B \rangle = 1.0$, $m_A = 0$ (Spatial model in the presence of migration of mutants only).

We anticipate that our results can be applied to various biological problems in cancer progression, bacterial habitats, embryogenesis and also in social and ecological modelling. As an example, we would like to explore the application of our results to cancer biology. It is well known that a tumour has a very complex microenvironment (independent of stage of the cancer). This microenvironment can have some adverse affects on the fitness and migration rates of host and mutant cells in the system, which in turn can impact the invasion probability. Moreover during carcinogenesis, it is quite possible that just two mutant phenotypes, with different fitness rates, exist at the early stages. We captured varying fitness rates at each point in the system by choosing the fitness matrix from different types of distributions such as uniform, triangular, bimodal. Our results suggest that the invasion probability of mutants into the proximal tissue depends on the fitness distribution, which is dictated by the microenvironmental conditions. Biologically, it is a known fact that several environmental stresses can lead to the epithelial-mesenchymal-transition (EMT) [41], which is an important component of metastasis (i.e. cancerous cells move and colonize a different organ). From our analysis we have shown that under certain conditions, a neutral mutant can turn into an advantageous mutant (in the absence of migration), and that migration amplifies the invasion probability even in heterogeneous conditions.

Our results can be used to understand the effects of heterogeneous microenvironments on phenotypic heterogeneity. We emphasize that our cellular automata rules are based on existing biological assumptions within a microenvironment. We have developed a computational framework to understand and analyze the importance of random fitness distributions on the invasion mechanism of mutants. Our results can be validated through biological experimental studies, and also to gain further insights into the effect of random fitnesses on the invasion probability. Although our computational model reflects a realistic heterogeneous scenario within a microenvironment, we can always refine the model to make a stronger clinical connection, by incorporating the fitness of a cell at a point on the grid as a function of various nutrients and oxygen supply that change temporally. Our model can be further extended to realistic tissue architectures through the study of invasion probability on unstructured meshes.

**Acknowledgments**

M. Kohandel and S. Sivaloganathan are supported by the Natural Sciences and Engineering Research Council of Canada (NSERC, discovery grants) as well as an NSERC/CIHR Collaborative Health Research grant.

**Author Contributions**

Performed the numerical experiments: VSKM; Analyzed data: VSKM/KK MK SS; Wrote the paper: VSKM/KK MK SS; Designed the project: KK VSKM MK SS.

**References**

1. Nowak MA. Evolutionary Dynamics: Exploring the Equations of Life, Belknap Press; 1 edition, 2006.

2. Broom M. Game-Theoretical Models in Biology, Chapman and Hall/CRC (2013).

3. Antal T, Redner S, Sood V. Evolutionary Dynamics on Degree-Heterogeneous Graphs. Phys Rev Lett. 2006;96(18):188104.

4. Sood V, Antal T, Redner S. Voter models on heterogeneous networks. Phys Rev E Stat Nonlin Soft Matter Phys. 2008;77(4 Pt 1):041121.

5. Maruyama T. A simple proof that certain quantities are independent of the geographical structure of population. Theor Popul Biol. 1974;5(2):148-54.

6. Maruyama T. A Markov Process of Gene Frequency Change in a Geographically Structured Population. Genetics. 1974;76(2):367-77.

7. Maruyama T. On the fixation probability of mutant genes in a sub-divided population. Genet Res. 1970;15(2):221-5.

8. Lieberman E, Hauert C, Nowak MA. Evolutionary dynamics on graphs. Nature.

2005;433(7023):312-6

9. Manem VSK, Kohandel M, Komarova NL, Sivaloganathan S. Spatial invasion dynamics on random and unstructured meshes: Implications for heterogeneous tumor populations. J Theor Biol. 2014;349:66-73.

10. Houchmandzadeh B, Vallade M. The fixation probability of a beneficial mutation in a geographically structured population, New Journal of Physics 13.7 (2011): 073020.

11. Masuda N, Gibert N, Redner S. Heterogeneous voter models, Phys Rev E Stat Nonlin Soft Matter Phys. 2010;82(1 Pt 1):010103.

12. Thalhauser CJ, Lowengrub JS, Stupack D, Komarova NL. Selection in spatial stochastic models of cancer: migration as a key modulator of fitness. Biol Direct. 2010;5:21.

13. Kamran K, Komarova NL, Kohandel M. The duality of spatial death–birth and birth–death processes and limitations of the isothermal theorem. R Soc Open Sci. 2015;2(4):140465.

14. Kepler TB, Perelson AS. Drug concentration heterogeneity facilitates the evolution of drug resistance. Proc Natl Acad Sci U S A. 1998;95(20):11514-9.

15. Moreno-Gamez S, Hill AL, Rosenbloom DI, Petrov DA, Nowak MA, Pennings PS. Imperfect drug penetration leads to spatial monotherapy and rapid evolution of multidrug resistance. Proc Natl Acad Sci U S A. 2015;112(22):E2874-83.

16. Feng F, Nowak MA, Bonhoeffer S. Spatial heterogeneity in drug concentrations can facilitate the emergence of resistance to cancer therapy. PLoS Comput Biol. 2015;11(3):e1004142.

17. Gevertz JL, Torquato S. Growing heterogeneous tumors in silico. Phys Rev E Stat Nonlin Soft Matter Phys. 2009;80(5 Pt 1):051910

18. Anderson ARA, Weaver AM, Cummings PT, Quaranta V. Tumor morphology and phenotypic evolution driven by selective pressure from microenvironment. Cell. 2006;127(5):905-15.

19. Gatenby RA, Vincent TL. Application of quantitative models from population biology and evolutionary game theory to tumor therapeutic strategies. Mol Cancer Ther. 2003;2(9):919-27.

20. Gatenby RA, Vincent TL. An evolutionary model of carcinogenesis. Cancer Res. 2003;63(19):6212-20.

21. Gatenby RA, Smallbone K, Maini PK, Rose F, Averill J, Nagle RB, et al. Cellular adaptations to hypoxia and acidosis during somatic evolution of breast cancer, Br J Cancer. 2007;97(5):646-53.

22. Gatenby RA, Gillies RJ. A microenvironmental model of carcinogenesis, Nat Rev Cancer. 2008;8(1):56-61. Review.

23. Whitlock MC and Gomulkiewicz R. Probability of Fixation in a Heterogeneous Environment. Genetics. 2005;171(3): 1407–1417.

24. Gavrilets, Sergey, Nathan Gibson. Fixation probabilities in a spatially heterogeneous environment. Population Ecology, 2002; 44(2): 51-58.

25. Hauser, Oliver P, Arne Traulsen, Martin A. Nowak. Heterogeneity in background fitness acts as a suppressor of selection. J Theor Biol. 2014;343:178-85

20. Deutsch A, Dormann S. Cellular automation modeling of biological pattern formation Birkhauser. Boston

26. Byrne H, Alarcón T, Owen M, Webb S, Maini P. Modeling Aspects of Cancer Dynamics: A Review. Philos Trans A Math Phys Eng Sci. 2006;364(1843):1563-78

27. Anderson A, Chaplain M, Rejniak K, Fozard J. Single-cell based models in biology and medicine. Math Med Biol, 25:185-186.

28. Anderson A, Quaranta V. Integrative mathematical oncology. Nat Rev Cancer. 2008;8(3):227-34.

29. Quaranta V, Rejniak K, Gerlee P, Anderson A. Invasion emerges from cancer cell adaptation to competitive microenvironments: Quantitative predictions from multiscale mathematical models. Semin Cancer Biol. 2008;18(5):338-48

30. Moran P. The Statistical Processes of Evolutionary Theory Oxford: Clarendon; 1962.

31. Komarova NL, Sengupta A, Nowak MA. Mutation-selection networks of cancer initiation: tumor suppressor genes and chromosomal instability. J Theor Biol. 2003;223(4):433-50.

32. Nowak MA, Michor F, Komarova NL, Iwasa Y. Evolutionary dynamics of tumor suppressor gene inactivation. Proc Natl Acad Sci U S A. 2004;101(29):10635-8.

33. Michor F, Iwasa Y, Rajagopalan H, Lengauer C, Nowak MA. Linear model of colon cancer initiation. Cell Cycle. 2004;3(3):358-62

34. Iwasa Y, Michor F, Nowak MA. Stochastic tunnels in evolutionary dynamics, Genetics. 2004;166(3):1571-9.

35. Komarova NL. Spatial stochastic models for cancer initiation and progression. Bull Math Biol. 2006;68(7):1573-99.

36. Komarova NL. Loss- and gain-of-function mutations in cancer: mass action, spatial and hierarchical models. Jour Stat Phys. 2007;128.1-2,413-446.

37. Foo J, Leder K, Michor F. Stochastic dynamics of cancer initiation. Phys Biol. 2011;8(1):015002

38. Fedotov S, Iomin A. Probabilistic approach to a proliferation and migration dichotomy in

tumor cell invasion. Phys Rev E Stat Nonlin Soft Matter Phys. 2008;77(3 Pt 1):031911.

39. Dykstra B, Ramunas J, Kent D, McCaffrey L, Szumsky E, Kelly L, Farn K, Blaylock A, Eaves C, Jervis E. High-resolution video monitoring of hematopoietic stem cells cultured in single-cell arrays identifies new features of self-renewal. Proc Natl Acad Sci U S A. 2006;103(21):8185-90.

40. Kimura M. Process leading to quasi-fixation of genes in natural populations due to random fluctuation of selection intensities. Genetics. 1954;39(3):280-95.

41. Weinberg RA. Biology of Cancer, Garland Science (2013).

**Figure captions**

**Figure 1:** Different forms of random fitness distributions: a) Uniform b) triangular c) bimodal.

**Figure 2:** Invasion probability of mutants as a function of varying distribution width of fitness $r_B$. Parameters: Lattice size=$21 \times 21$, $r_A = 1.0$, $\langle r_B \rangle = 1.5$, $m_A = m_B = 0$ (Spatial model in the absence of migration of cells).

**Figure 3:** Invasion probability of mutants as a function of varying distribution width of fitness $r_B$. Parameters: Lattice size=$21 \times 21$, $r_A = 1.0$, $\langle r_B \rangle = 1.0$, $m_A = m_B = 0$(Spatial model in the absence of migration of cells).

**Figure 4:** Invasion probability of mutants as a function of varying distribution width of fitness $r_B$. Parameters: Lattice size=$21 \times 21$, $\langle r_A \rangle = 1.0$ $\langle r_B \rangle = 1.5$, $m_A = m_B = 0$(Spatial model in the absence of migration of cells).

**Figure 5:** Invasion probability of mutants as a function of varying distribution width of fitness $r_B$. Parameters: Lattice size=$21 \times 21$, $\langle r_A \rangle = 1.0$ $\langle r_B \rangle = 1.0$, $m_A = m_B = 0$(Spatial model in the absence of migration of cells).

**Figure 6:** Invasion Probability against the percentage of advantageous mutants. Parameters: Lattice size=$21 \times 21$, $r_A = 1.0$, $m_A = m_B = 0$(Spatial model in the absence of migration of cells).

**Figure 7:** (a) Random distribution of two types of sites with high and low fitness's for mutant cells. (b) Evenly distributed high and low fitness sites with equal ratios (ideal case). This is the onset of successful fixation.

**Figure 8:** Comparison of invasion probabilities- Analytical and Simulated values for bimodal distributions of variable width (Spatial model in the absence of migration of cells).

**Figure 9:** Invasion probability of mutants as a function of varying distribution width of fitness $r_B$. Parameters: Lattice size=$21 \times 21$, $r_A = 1.0$, $\langle r_B \rangle = 1.5$, $m_A = 0$(Spatial model in the presence of migration of mutants only).

**Figure 10:** In the presence of migration the previously isolated regions with advantaged fitness for B cells can communicate due to their migration potential (indicated with arrows).

**Figure 11:** Invasion probability of mutants as a function of varying distribution width of fitness $r_B$. Parameters: Lattice size=$21 \times 21$, $r_A = 1.0$, $\langle r_B \rangle = 1.0$, $m_A = 0$(Spatial model in the presence of migration of mutants only).